# Critical Transitions in Public Opinion: A Case Study of American Presidential Election

Ning Ning Chung[1,2], Lock Yue Chew[1,2*] and Choy Heng Lai[3,‡]

**Affiliations:** [1]Complexity Institute, Nanyang Technological University Singapore 637973

[2]School of Physical & Mathematical Sciences, Nanyang Technological University Singapore 637377

[3]Department of Physics, National University of Singapore, Singapore 117551

Correspondence to: [*]lockyue@ntu.edu.sg and [‡]phylaich@nus.edu.sg

**Abstract**:

At the tipping point, it is known that small incident can trigger dramatic societal shift. Getting early-warning signals for such changes are valuable to avoid detrimental outcomes such as riots or collapses of nations. However, it is notoriously hard to capture the processes of such transitions in the real-world. Here, we demonstrate the occurrence of a major shift in public opinion in the form of political support. Instead of simple swapping of ruling parties, we study the regime shift of a party's popularity based on its attractiveness by examining the American presidential elections during 1980-2012. A single irreversible transition is detected in 1991. Once a transition happens, recovery to the original level of attractiveness does not bring popularity of the political party back. Remarkably, this transition is corroborated by tell-tale early-warning signature of critical slowing down. Our approach is applicable to shifts in public attitude within any social system.

**Main Text:**

Complex systems are known to undergo critical changes (1-8). Classic examples in this respect are natural ecosystems, where collapse has been observed in the fisheries stocks, in global climate change, and the bleaching of the coral reefs. It is usual for such critical transition to be triggered by a gradual perturbation or a seemingly minor happenstance. Once the transition begins, it can become drastic with an impact that is abrupt and irreversible. Interestingly, theory which governs ecosystem regime shifts has wide applicability. It has recently been applied even to the banking system (9-11). Shifts between alternate stable states have also been shown theoretically in social (12) and coupled socio-ecological systems (13). More empirical evidences are however needed to validate their existence in these systems. It is in these contexts that critical transition is demonstrated in this report in social systems of general interest. Here, we show how a model of opinion dynamics is directly applicable to recent American presidential elections with critical transitions predicted theoretically and duly verified empirically.

Opinions are dynamic and are heavily influenced by the underlying social interaction mechanisms. A notable mechanism is peers' influence, where individuals tend to follow the social opinion of their peers (14, 15). In consequence, they are more likely to adopt the opinion of the majority (16). There is also the situation where an individual persists on its opinion regardless of the views of its social environment. Such individuals are known as zealots (17, 18). However, few studies have been conducted on the intrinsic capacity of an opinion in affecting the choice of an individual. In this case, the opinion is deemed to bear a set of basic attributes which characterizes its quality and renders it appealing for adoption. We term this the *attractiveness* of the opinion.

Here, we introduce a theoretical model on two competing opinions $R$ and $D$. The model incorporates all the three mechanisms discussed above: majority rule; inertial to make social change; and implicit attractiveness of the opinion. Let us represent the agents of our social system by a network of size $N$ with average degree $k$. A fraction $f_R$ ($f_D$) of the $N$ agents are selected randomly to stick firmly to opinion $R$ ($D$) with $f_R + f_D \leq 1$, and these agents are zealots. The rest of the agents are flexible and they select their opinions by making their choice either through the majority-rule or the attractiveness of the opinion. At each time step, a flexible agent is randomly chosen to update its opinion. There is a probability $p_m$ for the selected agent to decide based on the majority-rule. In this case, the agent adopts the opinion followed by the majority of its peers. Conversely, the agent has a probability of $1 - p_m$ to decide with respect to the opinion's attractiveness. When this happens, it adopts opinion $R$ with probability $A_R$ and opinion $D$ with probability $A_D$ such that $A_R + A_D = 1$. We start our investigation on the system dynamics with a large $A_R$. The system is allowed to evolve for a sufficiently long time until it reaches steady state. After which, the steady-state value for the fraction of agents that adopt opinion $R$, i.e. $P_R$, is recorded. The value of $A_R$ is then gradually decreased. For each decrement of $A_R$, we record $P_R$. The process is then reversed by gradually increasing $A_R$ to its original value. This would give us the dynamical changes of opinion $R$ as its attractiveness varies from large to small and then back to its original value.

Depending on the value of $p_m$, $A_R$, $f_R$ and $f_D$, different opinion dynamics as exhibited by the $P_R$-$A_R$ plot is observed (see Fig. 2). When $p_m = 0$, $P_R$ and $A_R$ follows a monotically increasing one-to-one linear relationship. However, this functional behavior becomes nonlinear as $p_m$ is increased, where we observe the occurrence of a sharp transition between opinions. Beyond a critical $p_m$, the transition transforms further. We observe the interesting phenomenon of hysteresis and irreversible behaviour. The horizontal width of the hysteresis loop that emerges is found to increase as we continue to increase $p_m$. We expect $P_R = f_R$ when $A_R = 0$, and $P_R = 1 - f_D$ when $A_R = 1$.

Thus, a fundamental and unexpected outcome of the model is *critical* and *irreversible* transition of the dominance of the opinion from one to the other. It arises from the interplay between the mechanisms of majority rule and opinion's attractiveness. As the attractiveness of opinion $R$ ($A_R$) increases from the state where opinion $D$ is dominant (in the direction of single arrow in Fig 2e), the adoption of opinion $R$ increases. But this increase is restrained below the level possible from a monotonic linear increase in lieu that opinion $D$ is the majority. By symmetry, the same argument applies when we reduce $A_R$ (which is equivalent to an increase in opinion $D$'s attractiveness, $A_D$, and in the direction of the double arrow in Fig. 2e) from the state when $R$ is dominant. This explains why there is a sharp nonlinear change in between that accounts for the sudden switch of opinions. More significantly, this switch in opinion can occur at different $A_R$ in the forward and backward direction, leading to irreversible behavior and hysteresis loop.

While scientific investigation of presidential election has already been performed in various contexts (15, 19, 20), we hypothesize that our simple model possesses the main social mechanisms that drive the American presidential election. The American presidential election is held every four years via state elections. Each of the fifty states, plus Washington DC, is allocated with a certain number of electoral votes. In most states (except Maine and Nebraska), all allocated electoral votes are to be won through the majority votes. There are a total of 538 electoral votes and accomplishing a sum of 270 electoral votes (or more) in states' victory is all a president need to win an election. Interestingly, as we examine the American election results from 1980 to 2008 (21), we observe a ubiquitous and yet distinct voting pattern from states with different historical and social background. 13 of the states are found to be unfailingly Republican, while Minnesota and the districts of Columbia always vote for the Democrat during the period under study. The rest are swing states which sometimes vote for the Republican but other times for the Democrat.

We first examine the behavior of voters from the 13 Republican-committed states, which are shown in red in Fig. 1. A proxy for $A_R$ is defined based on the average vote for the Republican candidate for these 13 states (see Methods). Notably, we observe that the votes obtained by the Republican candidates in each of these states possess a simple increasing monotonic relationship with our surrogate $A_R$, for example, in Texas, Idaho and Wyoming (see Fig 2(a) − (c)). These results correspond to our model when $p_m$ is small and indicates that the average vote is a good measure of the attractiveness of a Republican candidate. Incidentally, this average vote reflects not only the attractiveness of $R$ in the Republican-committed states, it also implies the attractiveness of the Republican candidate in the Democrat-committed states, which is consistent with the outcome of our model. Votes obtained by the Republican in Minnesota, a Democratic state (shown in blue in Fig. 1), is found to increase linearly with the increase of $A_R$ as shown in Fig. 2(d).

In contrast, votes for swing states are found to depend nonlinearly on $A_R$. Let us take California as an example. While the Republican votes there may look random and unpredictable for a given attractiveness as shown in Fig. 2(e), the results become meaningful by following the history of the state election carefully. It reveals the interesting phenomenon of hysteresis as in the model. California was initially (1980-1988) a Republican-dominated state. The Republican votes in California were higher in 1984 as $R$ is more attractive in that year than 1980 and 1988. The Republican lost a large number of its followers as its attractiveness decreases over the four-year period when George Bush was the president of America. By the 1992 election, the Republican had become very unattractive and had a very low following in California. The critical transition from being popular to unpopular is observed to be hysteretic and irreversible. Eight years later in the 2000 election, the Republican had managed to regain its popularity in Texas and a few others, but not in California. Note that such an irreversible transition is also observed in 17 other swing states shown in green in Fig. 1. It is however not observed in the states which are colored yellow in Fig. 1.

It is significant that the above empirical observations correspond to the $P_R$ - $A_R$ relationship for each state obtained from our model. This implies that we can match theory to data through calibrating the parameters $p_m$, $f_R$ and $f_D$. The results are illustrated in Fig. 2. Good correspondence is found for all the states (see Supplementary Material for the rest of the states).

Critical slowing down precedes critical transition. It gives the most important clue (22-26) and early warning that the system is getting close to the transition point. During the period of our study from 1980 to 2008, we detect its symptom (see Fig. 3) in 1991, which is one year before the 1992 election, through the presidential approval rating (PAR). Note that political analysts (27) have identified PAR as a good global indicator on election outcome. The result here is consistent with our observation above, as critical slowing down is found to occur just prior to the period of critical irreversible transition in the support of the Republican during the tenure of George Bush. When a system is close to the transition point, the rate of recovering from small perturbation is lower, and its dynamics is characterized by high correlation and large standard deviation as shown in Fig. 3(b).

Let us next test our model by using it to predict the 2012 election. In order to make prediction, our model requires a measure on the perception of the candidate at the population level. For this, we shall use PAR in conjunction with a new quantity $\langle P_R \rangle$ for our prediction. $\langle P_R \rangle$ ($\langle P_D \rangle$) is the average fraction of votes for the 51 electoral colleges cast in favor of the Republican (Democratic) candidate for each election year. The appropriateness of $\langle P_R \rangle$ is discerned from its close correspondence to the PAR (except when there was a change in the incumbent political party in the year 2000 and 2008 elections, see Fig. 3). Specifically, we observe that $\langle P_R \rangle \approx$ PAR when the president is a Republican, and $\langle P_R \rangle \approx 1$ - PAR when the president is a Democrat since $\langle P_R \rangle = 1 - \langle P_D \rangle$. We can easily compute the relationship between $\langle P_R \rangle$ and $A_R$ by making use of the 51 $P_R$ - $A_R$ relations that was determined earlier. The $\langle P_R \rangle$ - $A_R$ relation is necessary to predict the outcome of future presidential election.

For the 2012 election, we use 1 - PAR = 49.9 (with 50.1 being the PAR on October 2012 of President Obama) to make prediction. Among the 51 electoral colleges, election results for three of the states, namely Florida, Virginia and Colorado are found to be the least predictable (see Fig. S7) because the statistical error of their $P_R$ is larger than their winning margin. Notably, these three states are among the seven super-swingy states identified by Sabato et. al. in their 2016 electoral map (28). Our predictions for the Republican votes from our model are shown in Table 1. On average, we observe a discrepancy of 2.7% between the predicted and the actual votes. This has led to a difference of 13 electoral votes between the results of our prediction (219 EV) and those based on the actual election (206 EV).

The popularity level between the Democratic candidate Hillary Clinton and the Republican candidate Donald Trump has been tracked by RealClearPolitics (29) since May 2015. While the favorability rating of Clinton is way ahead of Trump initially, their gap has finally closed recently after much fluctuations, with Clinton slightly ahead with a spread of 4. Interestingly, president Obama's approval rating has remain at around 52 according to opinion polls performed by Gallup in Sep 2016 (30). If this rating persists till November 2016, the Republican could end up losing the election with an electoral vote that ranges between 175 and 228 (see Table S1). On the other hand, competition between the two parties can become far more intense if the Republican candidate is able to acquire a $\langle P_R \rangle$ of 51.7 which would enable them to win the election. In this case, the votes for the two parties would be extremely close in the 11 states marked red in Table S1.

**Methods**

<u>Proxy for Attractiveness</u>

The proxy for attractiveness $A_{Rj}$ is obtained based on the average vote for the Republican candidate for the 13 Republican-committed states $\bar{v}_{Rj}$ of the $j$-th election as follows:

$$A_{Rj} = \bar{v}_{Rj} - 0.5 \left\langle |\bar{v}_{Rj} - \bar{V}_{Rj}| \right\rangle .$$

In order for this proxy not to be too far off from the average vote over the 51 electoral states $\bar{V}_{Rj}$, we have subtracted a constant term $0.5 \left\langle |\bar{v}_{Rj} - \bar{V}_{Rj}| \right\rangle$ which represents the mean discrepancy between the two average votes over the eight election years.

<u>$L^2$ Error</u>

The $L^2$ error is defined as follows:

$$e_{L^2} = \sqrt{\sum_j (v_j{}^M - v_j{}^D)^2} .$$

Note that $v_j{}^M$ is the vote determined from our model, while $v_j{}^D$ is the actual vote obtained empirically. Both are for the $j$-th election.

Figure 1: Classification of US states based on voting patterns at state election level held between 1980 and 2008. States are categorized into Republican-committed states (red), Democratic-committed states (blue), swing states in which hysteresis are observed (green) and swing states in which hysteresis are not observed (yellow).

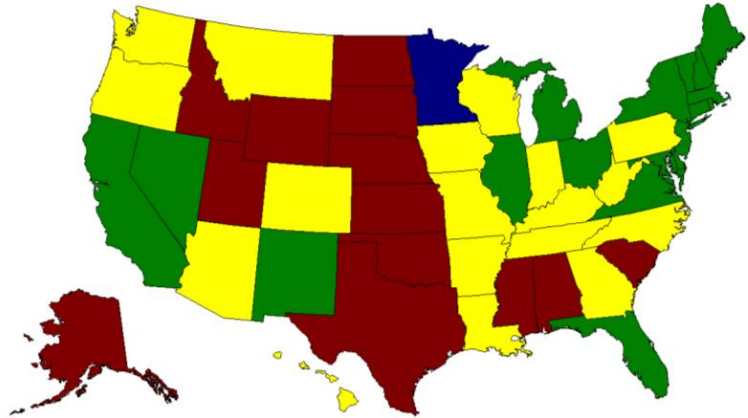

Figure 2: Dependence of the fraction of vote for the Republican candidate $P_R$ on its attractiveness $A_R$ for 8 American states. Empirical election results are shown as filled circles with the size of the circle increases with the year of the election. Results from our model are shown as solid line. Here, critical points for figures (e) to (h) are estimated from 40 simulations with standard deviations ranging from 0.02 to 0.04.

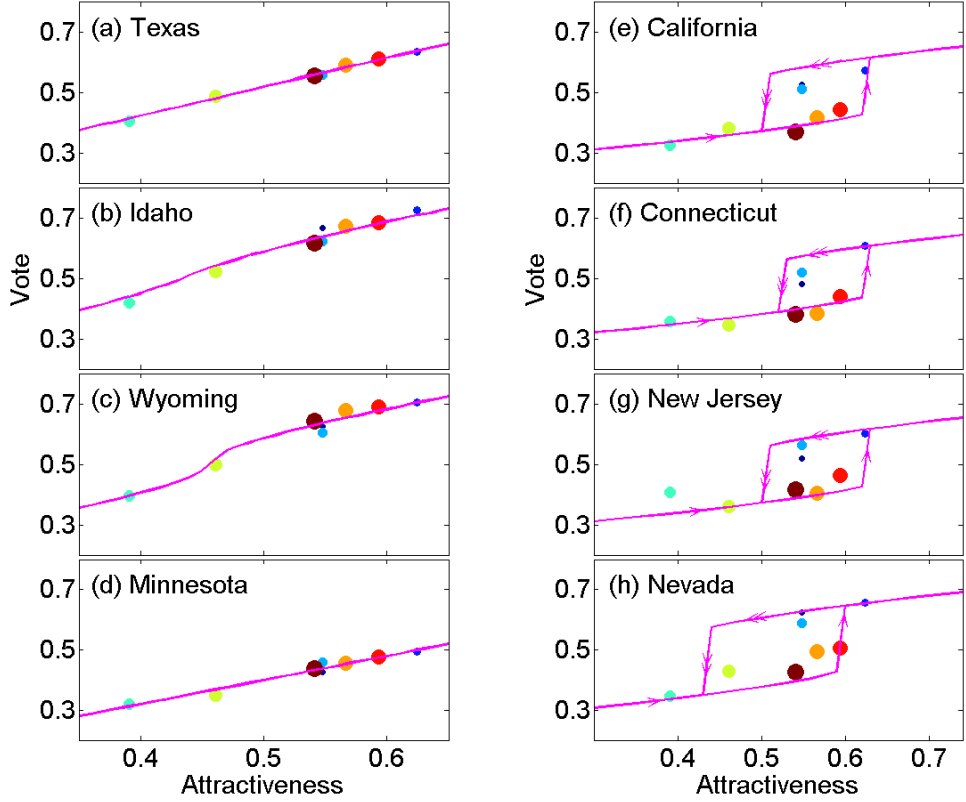

Figure 3: (a) The presidential approval rating (PAR) for American presidents from 1980 to 2016. Note that PAR are shown as it is for the Republican presidents (red curve), but for Democratic presidents, 1-PAR are shown instead (blue curve). The Republican votes (averaged over 51 electoral colleges), i.e. $\langle P_R \rangle$, are shown as circles for comparison. Autocorrelation lagged by seven days are shown for presidential approval rating between (b) Mar 1991 and Oct 1991, (c) Jan 1999 and Aug 1999, (d) Feb 2007 and Oct 2008, (e) Sep 2015 and Apr 2016. Critical slowing down is observed as an indicator that the system is near to the 'tipping point' in 1991.

(a)

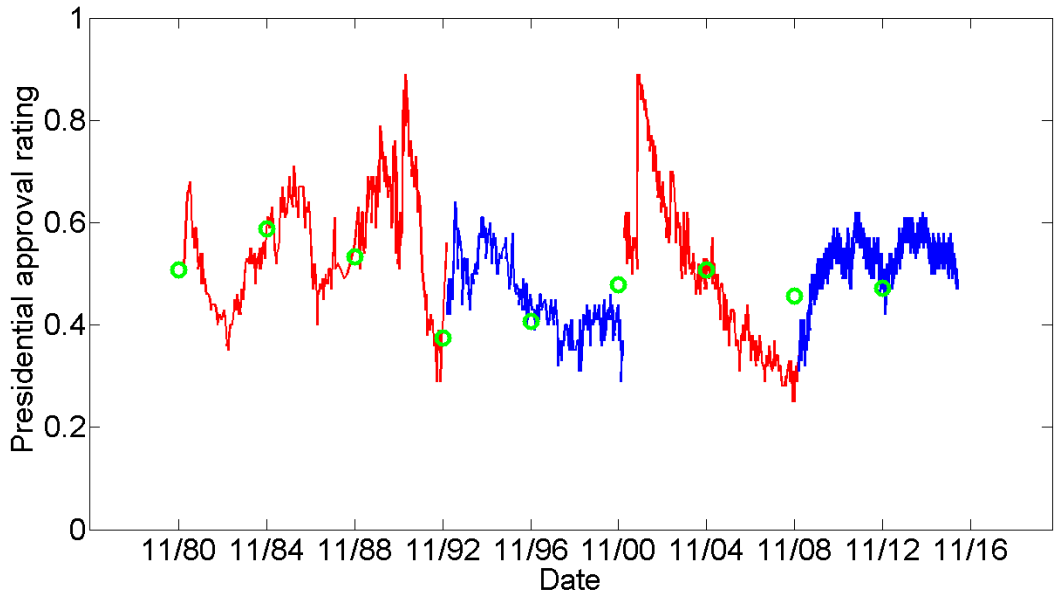

(b)

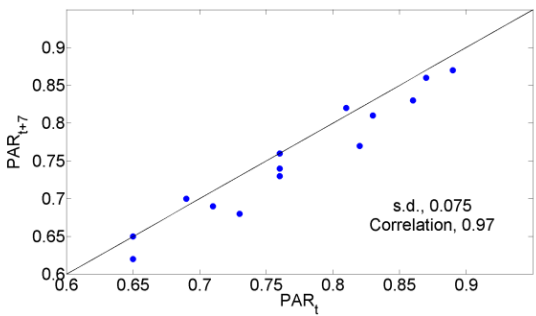

(c)

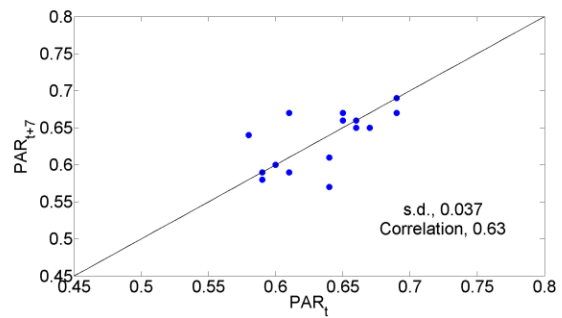

(d)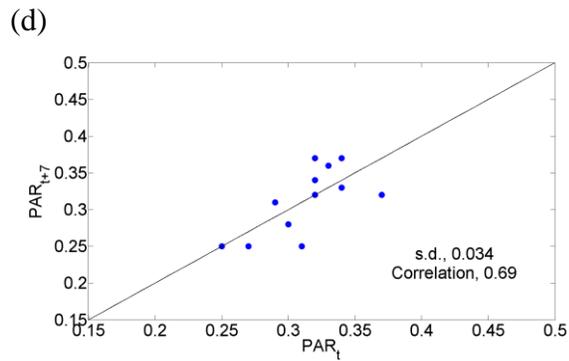     (e)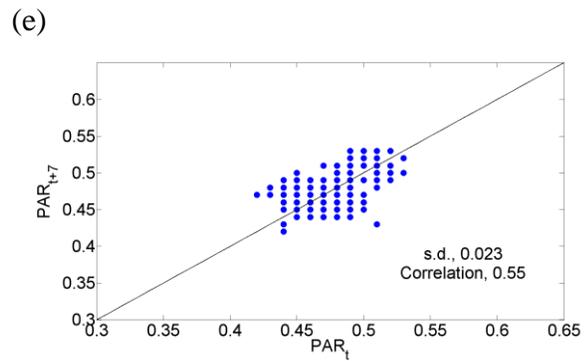

Table 1. Prediction of the election results based on the Oct 2012 Presidential Approval Rating of 1-PAR = 49.9. Note that the average vote for the Republican candidate in the 2012 election is 49.0.

| Electoral College | 2012 Election | | Model | | Discrepancy | |
|---|---|---|---|---|---|---|
| | Vote | EV | Vote | EV | Vote | EV |
| 1. Alabama | 60.5 | 9 | 57.6 | 9 | -2.9 | 0 |
| 2. Alaska | 54.8 | 3 | 60.0 | 3 | +5.2 | 0 |
| 3. Arizona | 53.7 | 11 | 56.4 | 11 | +2.7 | 0 |
| 4. Arkansas | 60.6 | 6 | 54.9 | 6 | -5.7 | 0 |
| 5. California | 37.1 | - | 40.6 | - | +3.5 | 0 |
| 6. Colorado | 46.1 | - | 48.9 | - | +2.8 | 0 |
| 7. Connecticut | 40.7 | - | 43.1 | - | +2.4 | 0 |
| 8. Delaware | 40.0 | - | 42.4 | - | +2.4 | 0 |
| 9. Dist. Of Col. | 7.3 | - | 10.7 | - | +3.4 | 0 |
| 10. Florida | 49.1 | - | 48.5 | - | -0.6 | 0 |
| 11. Georgia | 53.3 | 16 | 55.1 | 16 | +1.8 | 0 |
| 12. Hawaii | 27.8 | - | 37.3 | - | +9.5 | 0 |
| 13. Idaho | 64.5 | 4 | 65.6 | 4 | +1.1 | 0 |
| 14. Illinois | 40.7 | - | 40.5 | - | -0.2 | 0 |
| 15. Indiana | 54.1 | 11 | 56.9 | 11 | +2.7 | 0 |
| 16. Iowa | 46.2 | - | 48.4 | - | +2.2 | 0 |
| 17. Kansas | 59.7 | 6 | 59.4 | 6 | -0.3 | 0 |
| 18. Kentucky | 60.5 | 8 | 58.0 | 8 | -2.5 | 0 |
| 19. Louisiana | 57.8 | 8 | 55.2 | 8 | -2.6 | 0 |
| 20. Maine | 41.0 | - | 42.4 | - | +1.4 | 0 |
| 21. Maryland | 35.9 | - | 40.0 | - | +4.1 | 0 |
| 22. Massachusetts | 37.5 | - | 33.7 | - | -3.8 | 0 |
| 23. Michigan | 44.7 | - | 47.5 | - | +2.8 | 0 |
| 24. Minnesota | 45.0 | - | 45.6 | - | +0.6 | 0 |
| 25. Mississippi | 55.3 | 6 | 57.0 | 6 | +1.7 | 0 |
| 26. Missouri | 53.8 | 10 | 52.4 | 10 | -1.4 | 0 |
| 27. Montana | 55.4 | 3 | 55.8 | 3 | +0.4 | 0 |
| 28. Nebraska | 59.8 | 5 | 63.9 | 5 | +4.1 | 0 |
| 29. Nevada | 45.7 | - | 44.6 | - | -1.1 | 0 |
| 30. New Hampshire | 46.5 | - | 46.8 | - | +0.3 | 0 |
| 31. New Jersey | 40.6 | - | 40.6 | - | 0.0 | 0 |
| 32. New Mexico | 42.8 | - | 44.9 | - | +2.1 | 0 |
| 33. New York | 35.2 | - | 38.1 | - | +2.9 | 0 |
| 34. North Carolina | 50.4 | 15 | 53.7 | 15 | +3.3 | 0 |
| 35. North Dakota | 58.3 | 3 | 60.7 | 3 | +2.4 | 0 |
| 36. Ohio | 47.7 | - | 46.8 | - | -0.9 | 0 |
| 37. Oklahoma | 66.8 | 7 | 62.5 | 7 | -4.3 | 0 |
| 38. Oregon | 42.1 | - | 45.6 | - | +3.5 | 0 |
| 39. Pennsylvania | 46.7 | - | 47.9 | - | +1.2 | 0 |
| 40. Rhode Island | 35.2 | - | 33.7 | - | -1.5 | 0 |
| 41. South Carolina | 54.6 | 9 | 57.3 | 9 | +2.7 | 0 |
| 42. South Dakota | 57.9 | 3 | 58.1 | 3 | +0.2 | 0 |
| 43. Tennessee | 59.5 | 11 | 55.6 | 11 | -3.9 | 0 |
| 44. Texas | 57.2 | 38 | 58.6 | 38 | +1.4 | 0 |
| 45. Utah | 72.8 | 6 | 68.3 | 6 | -4.5 | 0 |
| 46. Vermont | 31.0 | - | 35.6 | - | +4.6 | 0 |
| 47. Virginia | 47.3 | - | 51.7 | 13 | +4.4 | +13 |
| 48. Washington | 41.3 | - | 46.7 | - | +5.4 | 0 |
| 49. West Virginia | 62.3 | 5 | 55.6 | 5 | -6.7 | 0 |
| 50. Wisconsin | 46.0 | - | 48.4 | - | +2.4 | 0 |
| 51. Wyoming | 68.6 | 3 | 65.4 | 3 | -3.2 | 0 |
| | Total | 206 | | 219 | | 13 |

**References:**

1. Scheffer M, Carpenter S, Foley JA, Folke C, Walker B (2001) Catastrophic Shifts in Ecosystems. *Nature* 413: 591.

2. Scheffer M, Carpenter S, Lenton T, Bascompte J, Brock W, Dakos V, van de Koppel J, Leemput I, Levin S, van Nes E, Pascual M, Vandermeer J (2012) Anticipating Critical Transitions. *Science* 338(6105): 344-348.

3. Tripati A, Backman J, Elderfield H, Ferretti P (2005) Eocene bipolar glaciation associated with global carbon cycle changes. *Nature* 436: 341-346.

4. Liu Z, et al. (2009) Global cooling during the Eocene-Oligocene climate transition. *Science* 323: 1187-1190.

5. Barnosky AD, et al. (2012) Approaching a State Shift in Earth's Biosphere. *Nature* 486: 52.

6. Kefi S, Rietkerk M, Alados CL, Pueyo Y, Papanastasis, VP, Elaich A, Ruiter PC (2007) Spatial Vegetation Patterns and Imminent Desertification in Mediterranean Arid Ecosystems. *Nature* 449: 213.

7. Rietkerk M, Dekker SC, de Ruiter PC, van de Koppel J (2004) Self-organized patchiness and catastrophic shift in ecosystems. *Science* 305: 1926-1929.

8. Christianen MJA, et al. (2014) Habitat collapse due to overgrazing threatens turtle conservation in marine protected areas. *Proc R Soc B* 281: 2890.

9. May RM, Levin S, Sugihara G (2008) Ecology for Bankers. *Nature* 451: 893.

10. Haldane AG, May RM (2011) Systemic Risk in Banking Ecosystems. *Nature* 469: 351.

11. Kambhu J, Weidman S, Krishnan N (2007) New Directions for Understanding Systemic Risk (National Academies Press, Washington DC). Also published as Econ Policy Rev 13: 2.

12. Ohta M, Iida T, Kawaoka T (1995) Opinion Transition Model under Dynamic Environment: Experiment in Introducing Personality to Knowledge-based Systems. *Industrial and Engineering Applications of Artificial Intelligence and Expert Systems:* p. 569-574.

13. Lade SJ, Tavoni A, Levin S, Schlüter M (2013) Regime Shift in a Social-ecological System. *Theoretical Ecology* 6: 359-372.

14. Liggett TM (1985) *Interacting Particle System* (Springer, New York).

15. Fernández-Gracia J, Suchecki K, Ramasco JJ, Miguel MS, Eguíluz VM (2014) Is the Voter Model a Model for Voters? *Physical Review Letters* 112: 158701.

16. Galam S (2002) Minority Opinion Spreading in Random Geometry. *European physics Journal B* 25: 403-406.

17. Crokidakis N, Oliveira PMC (2015) Inflexibility and Independence: Phase transitions in the Majority-rule Model. *Physical Review E* 92: 062122.

18. Waagen A, Verma G, Chan K, Swami A, D'Souza RM (2015) Effect of Zealotry in High-dimensional Opinion Dynamics Models. *Physical Review E* 91: 022811.

19. Fortunato S, Castellano C (2007) Scaling and Universality in Proportional Elections. *Physical Review Letters* 99: 138701.

20. Klimek P, Yegorov Y, Hanel R, Thurner S (2012) Statistical Detection of Systematic Election Irregularities. *Proc Natl Acad Sci USA* 109: 16469-16473.

21. Source: The American Presidency Project.

22. van Nes EH, Scheffer M (2007) Slow recovery from perturbations as generic indicator of a nearby catastrophic shift. *Am Nat* 169(6): 738-747.

23. Scheffer M, et al. (2009) Early-warning signals for Critical Transitions. *Nature* 461(7260): 53-59.

24. Drake JM, Griffen BD (2010) Early Warning Signal of Extinction in Deteriorating Environments. *Nature* 467(7314): 456-459.

25. Dai L, Korolev KS, Gore J (2013) Slower Recovery in Space Before Collapse of Connected Populations. *Nature* 496: 355.

26. Leemput IA, et al. (2014) Critical Slowing Down as Early Warning for the Onset and Termination of Depression. *Proc Natl Acad Sci USA* 111: 87-92.

27. Kyle K, Geoffrey S, Sabato LJ (2015) Clinton's Real Opponent: Barack Obama. *Politico* April 17.

28. Kyle K, Geoffrey S, Sabato LJ (2015) The 2016 Results We Can Already Predict. *Politico* May 3.

29. Source: http://www.realclearpolitics.com

30. Source: http://www.gallup.com

**Supplementary Materials:**

Figure S1: Dependence of the fraction of vote for the Republican candidate $P_R$ on its attractiveness $A_R$. The relations are shown here for 8 states with hysteresis.

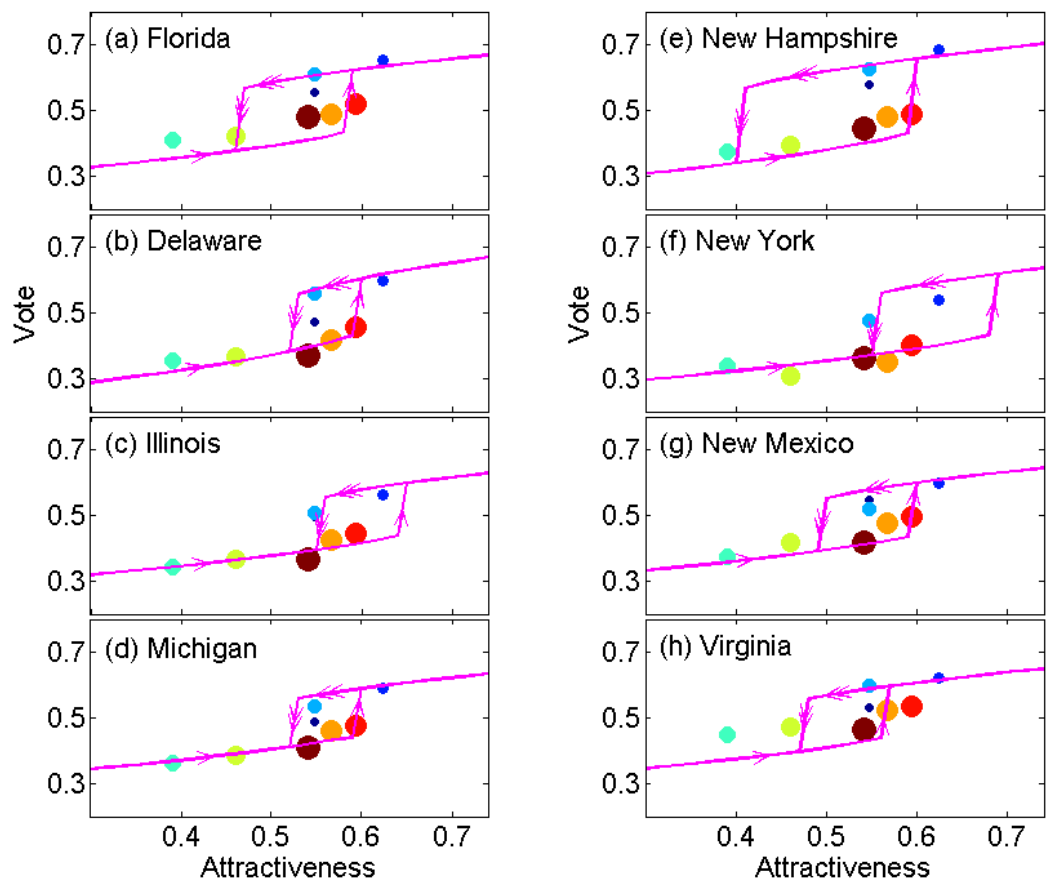

Figure S2: Dependence of the fraction of vote for the Republican candidate $P_R$ on its attractiveness $A_R$. The relations are shown here for 6 states with hysteresis.

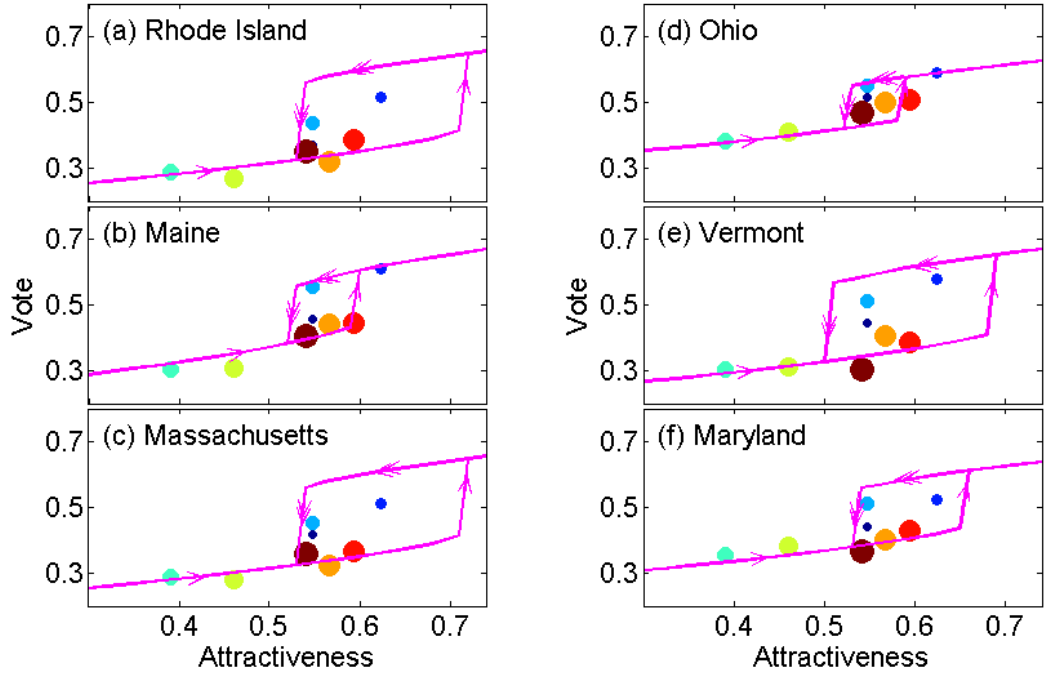

Figure S3: Dependence of the fraction of vote for the Republican candidate $P_R$ on its attractiveness $A_R$. The relations are shown here for 8 states without hysteresis.

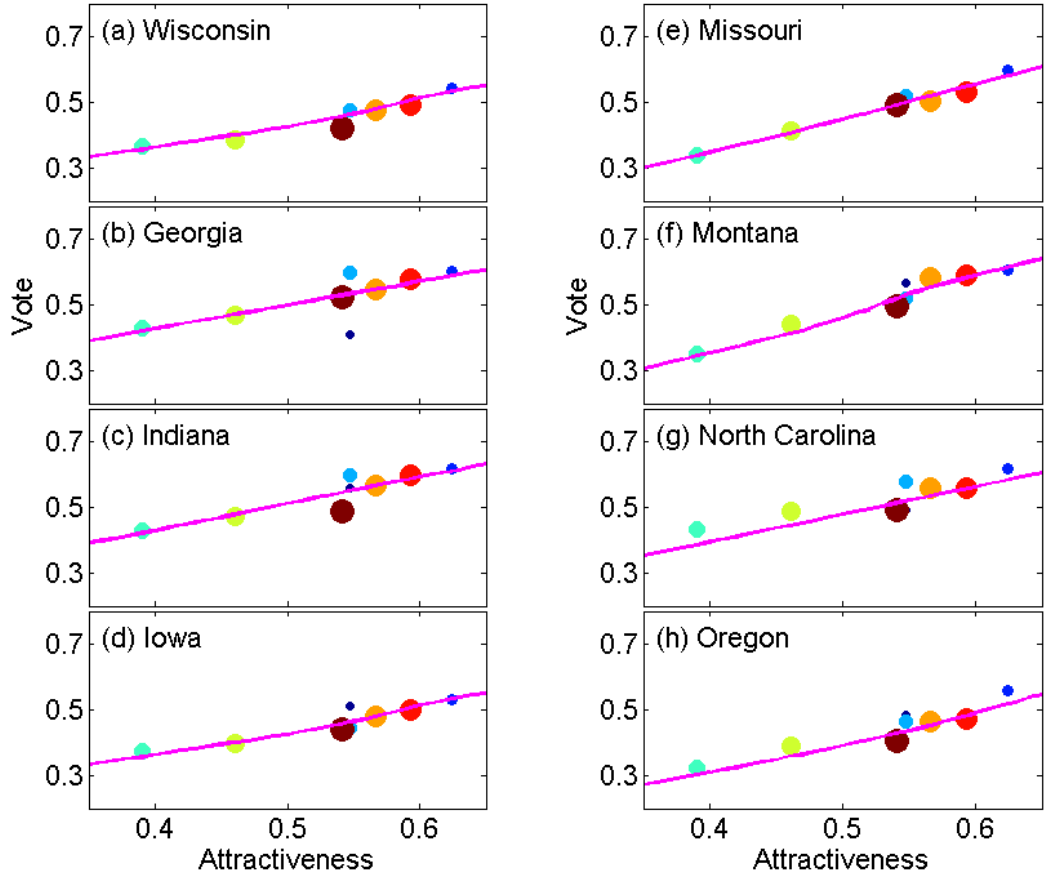

Figure S4: Dependence of the fraction of vote for the Republican candidate $P_R$ on its attractiveness $A_R$. The relations are shown here for 7 states without hysteresis.

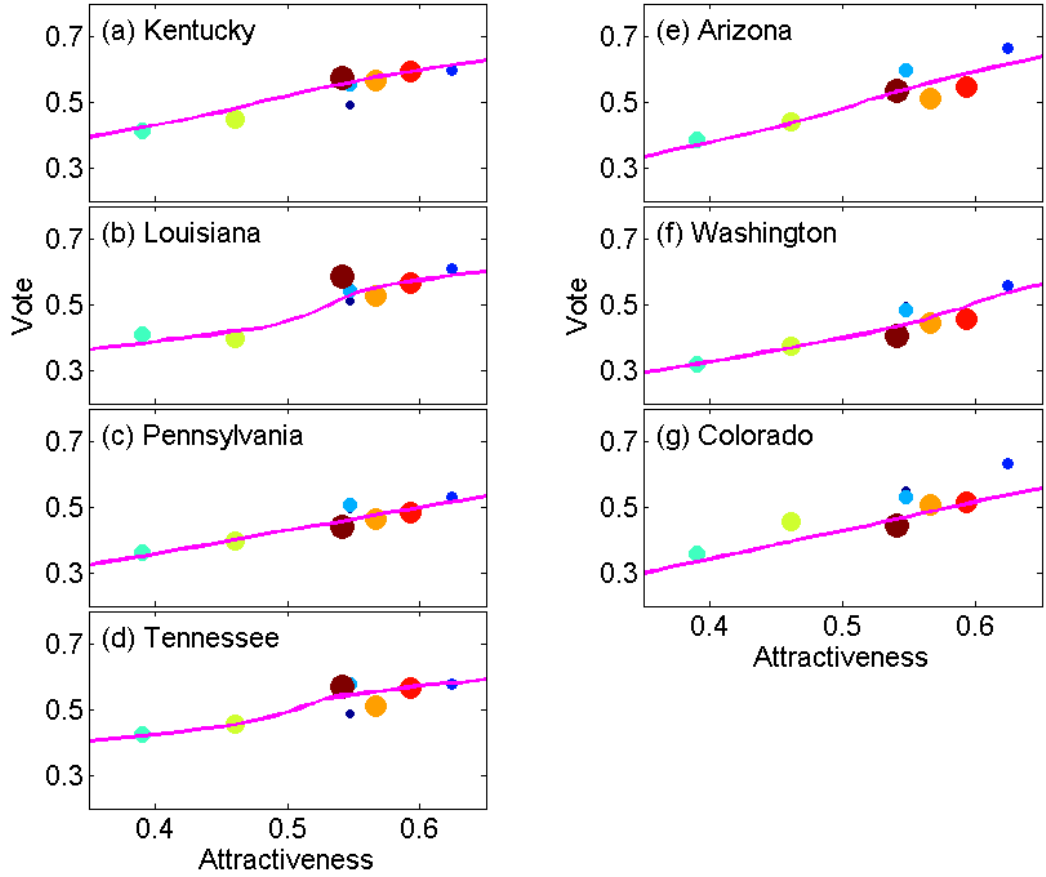

Figure S5: Dependence of the fraction of vote for the Republican candidate $P_R$ on its attractiveness $A_R$. The relations are shown here for 8 states without hysteresis.

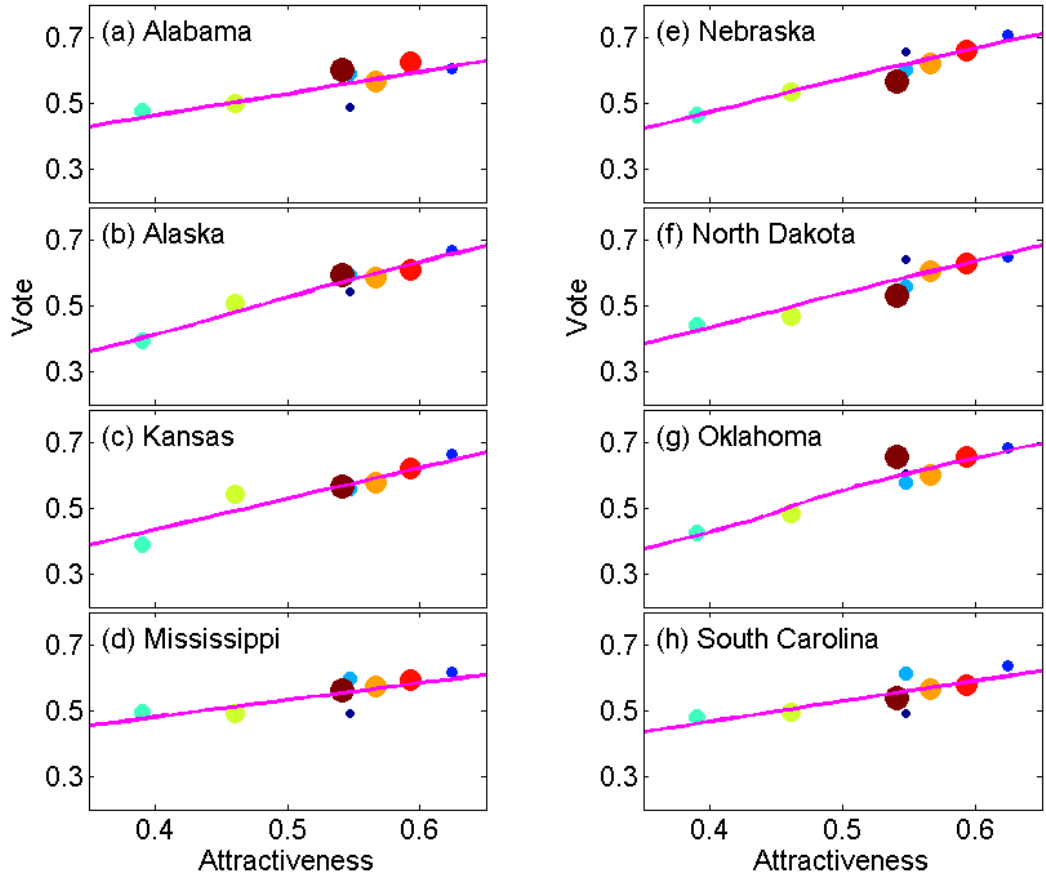

Figure S6: Dependence of the fraction of vote for the Republican candidate $P_R$ on its attractiveness $A_R$. The relations are shown here for 6 states without hysteresis.

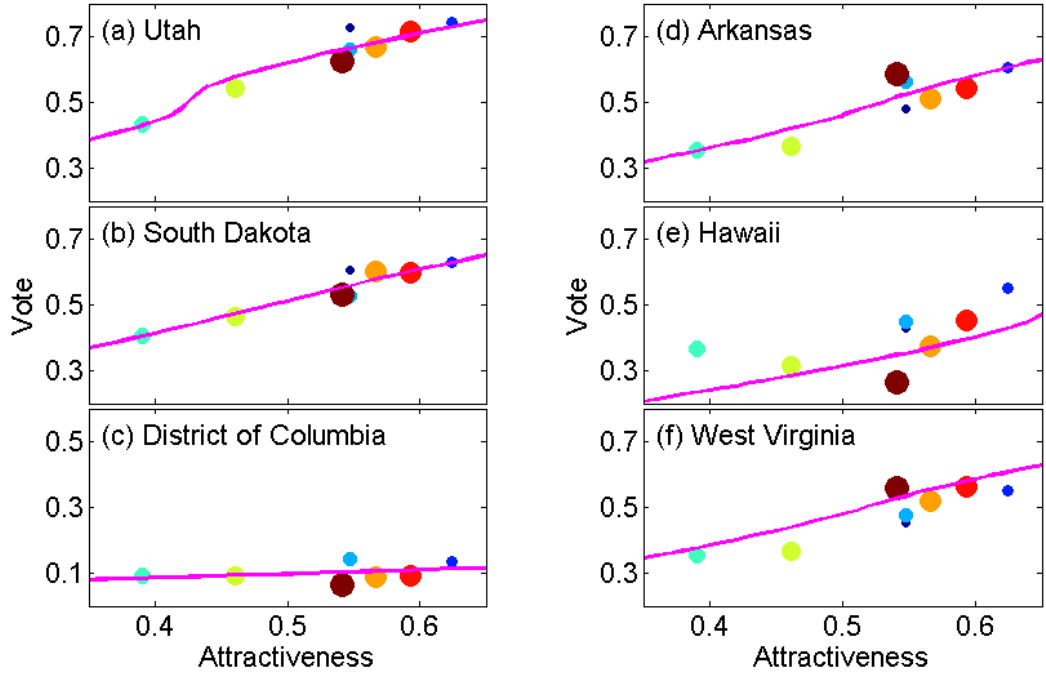

Figure S7: Statistical analysis performed on our model to determine its predictability on winning of the state election based on 1 - PAR = 49.9. In order to predict a win at the state level, the $L^2$ error of the estimate of the vote should be less than four times the winning margin $M$, where $M = |P_R - P_D|$. We have taken a stricter criterion of $L^2$ error = $2M$ as the boundary for predictability.

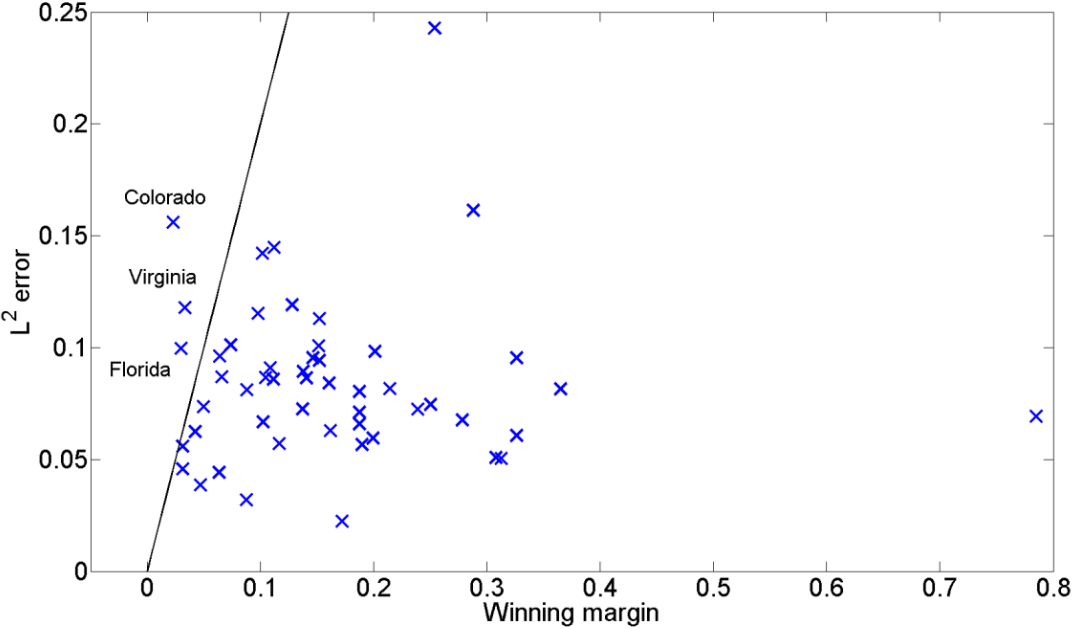

Figure S8: Further check on the validity of our model by examining the relationship between the number of acquired electoral votes and $\langle P_R \rangle$. Good correspondence is observed between model and data not only on the direct relationship between these variables, but also on the occurrence of historical precedence in their relationships, except for the election years: 1980, 1992, 1996 and 2000, where the presence of an independent candidate is not negligible. Note that the dashed line (solid line) represents votes obtained before (after) transition of opinions from our model. The circle marker denotes election results derived empirically. The electoral votes are allocated based on the 2010 Census which is effective for the 2012, 2016 and 2020 American presidential elections.

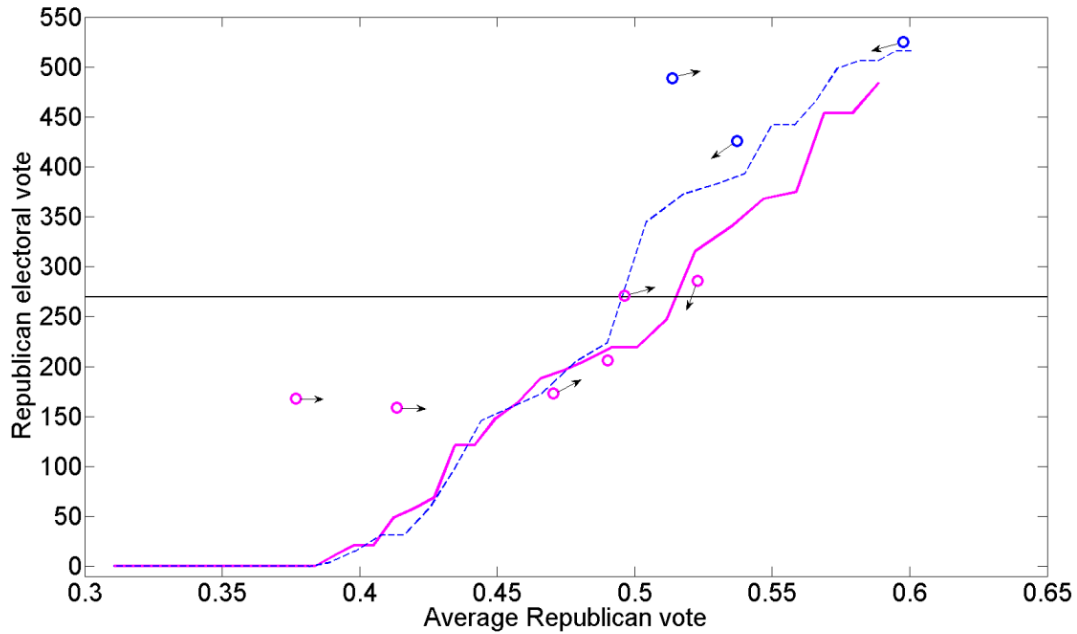

Table S1. Prediction of the 2016 election results based on an average vote for the Republican candidate $\langle P_R \rangle$ of 51.7 and 48.

| Electoral College | With average vote of 51.7 | | With average vote of 48 | |
|---|---|---|---|---|
| | Vote | EV | Vote | EV |
| 1. Alabama | 58.7 | 9 | 56.2 | 9 |
| 2. Alaska | 61.7 | 3 | 58.0 | 3 |
| 3. Arizona | 58.0 | 11 | 53.9 | 11 |
| 4. Arkansas | 56.8 | 6 | 52.6 | 6 |
| 5. California | 43.6 | - | 39.8 | - |
| 6. Colorado | 50.4 | 9 | 47.1 | - |
| 7. Connecticut | 44.1 | - | 41.1 | - |
| 8. Delaware | 46.8 | - | 40.1 | - |
| 9. Dist. Of Col. | 10.8 | - | 10.4 | - |
| 10. Florida | 51.9 | 29 | 44.2 | - |
| 11. Georgia | 56.1 | 16 | 53.5 | 16 |
| 12. Hawaii | 38.9 | - | 35.2 | - |
| 13. Idaho | 67.5 | 4 | 63.8 | 4 |
| 14. Illinois | 41.6 | - | 39.9 | - |
| 15. Indiana | 58.1 | 11 | 55.0 | 11 |
| 16. Iowa | 49.8 | - | 46.4 | - |
| 17. Kansas | 61.0 | 6 | 57.3 | 6 |
| 18. Kentucky | 58.8 | 8 | 56.1 | 8 |
| 19. Louisiana | 56.8 | 8 | 53.0 | 8 |
| 20. Maine | 46.6 | - | 40.1 | - |
| 21. Maryland | 40.8 | - | 39.0 | - |
| 22. Massachusetts | 34.4 | - | 32.9 | - |
| 23. Michigan | 50.2 | 16 | 44.3 | - |
| 24. Minnesota | 46.8 | - | 43.7 | - |
| 25. Mississippi | 57.8 | 6 | 55.9 | 6 |
| 26. Missouri | 54.3 | 10 | 49.9 | - |
| 27. Montana | 57.7 | 3 | 53.4 | 3 |
| 28. Nebraska | 65.4 | 5 | 62.1 | 5 |
| 29. Nevada | 47.9 | - | 41.6 | - |
| 30. New Hampshire | 51.5 | 4 | 42.3 | - |
| 31. New Jersey | 43.6 | - | 39.8 | - |
| 32. New Mexico | 49.0 | - | 42.4 | - |
| 33. New York | 39.0 | - | 37.4 | - |
| 34. North Carolina | 55.1 | 15 | 52.0 | 15 |
| 35. North Dakota | 62.2 | 3 | 58.7 | 3 |
| 36. Ohio | 49.9 | - | 45.3 | - |
| 37. Oklahoma | 64.1 | 7 | 60.6 | 7 |
| 38. Oregon | 47.4 | - | 43.4 | - |
| 39. Pennsylvania | 49.0 | - | 46.4 | - |
| 40. Rhode Island | 34.4 | - | 32.9 | - |
| 41. South Carolina | 58.2 | 9 | 56.1 | 9 |
| 42. South Dakota | 59.5 | 3 | 55.8 | 3 |
| 43. Tennessee | 56.7 | 11 | 54.8 | 11 |
| 44. Texas | 60.1 | 38 | 56.6 | 38 |
| 45. Utah | 69.9 | 6 | 66.5 | 6 |
| 46. Vermont | 36.4 | - | 34.8 | - |
| 47. Virginia | 55.0 | 13 | 47.8 | - |
| 48. Washington | 48.4 | - | 44.2 | - |
| 49. West Virginia | 57.2 | 5 | 53.8 | 5 |
| 50. Wisconsin | 49.8 | - | 46.4 | - |
| 51. Wyoming | 67.0 | 3 | 63.6 | 3 |
| Total | | 277 | | 196 |